\documentclass[12pt,letterpaper]{JHEP3}
\usepackage{amsmath,amssymb,amsfonts,cite}

\title{A Farey tale for $\CN=4$ dyons}

\preprint{}
\author{
Sameer Murthy and Boris Pioline \\
\it {Laboratoire de Physique Th\'eorique et Hautes Energies (LPTHE)\\
\it{Universit\'e Pierre et Marie Curie-Paris 6; CNRS UMR 7589}\\
\it{Tour 24-25, 5$^{\grave{e}me}$ \'etage, Boite 126, 4 Place Jussieu} \\
\it {75252 Paris Cedex 05, France}}\\
\email{[smurthy,pioline]@lpthe.jussieu.fr}
}

\abstract{We study exponentially suppressed contributions to the degeneracies of extremal
black holes. Within  Sen's quantum entropy function framework and focusing on   
extremal black holes with an intermediate $AdS_3$ region, we identify an infinite family of
semi-classical $AdS_2$ geometries which can contribute effects of order $\exp(S_0/c)$,
where $S_0$ is the Bekenstein-Hawking-Wald entropy and $c$ is an integer greater than
one.  These solutions lift to the extremal limit of the $SL(2,\IZ)$ family of 
BTZ black holes familiar from  the ``black hole Farey tail". 
We test this understanding in $\cN=4$ string vacua, where exact dyon degeneracies
are known to be given by Fourier coefficients of Siegel modular forms. We relate the sum
over poles in the Siegel upper half plane to the Farey tail expansion, and derive a
``Farey tale" expansion for the dyon partition function. Mathematically, this provides 
a (formal) lift from Hilbert modular forms to Siegel modular forms with a pole at the
diagonal divisor.}


\newcommand{\bea}{\begin{eqnarray}}
\newcommand{\eea}{\end{eqnarray}}
\newcommand{\be}{\begin{equation}}
\newcommand{\ee}{\end{equation}}

\newcommand{\de}{\mathrm{d}}

\newcommand{\I}{\mathrm{i}}

\newcommand{\p}{\partial}

\newcommand{\cA}{\mathcal{A}}
\newcommand{\CA}{\mathcal{A}}

\newcommand{\CC}{\mathcal{C}}

\newcommand{\cS}{\mathcal{S}}

\newcommand{\CL}{\mathcal{L}}

\newcommand{\cN}{\mathcal{N}}
\newcommand{\CN}{\mathcal{N}}
\newcommand{\cO}{\mathcal{O}}

\newcommand{\CZ}{\mathcal{Z}}

\renewcommand{\Im}{{\rm Im}}
\renewcommand{\Re}{{\rm Re}}

\newcommand{\pa}{\partial}
\newcommand{\nn}{\nonumber}
\newcommand{\IR}{\mathbb{R}}
\newcommand{\IC}{\mathbb{C}}
\newcommand{\IZ}{\mathbb{Z}}

\newcommand{\Pf}{\mbox{Pf}}

\def\bea{\begin{eqnarray}}
\def\eea{\end{eqnarray}}
\def\be{\begin{equation}}
\def\ee{\end{equation}}
\def\ba{\begin{align}}
\def\ea{\end{align}}
\def\bse{\begin{subequations}}
\def\ese{\end{subequations}}

\def\ellp{\ell}

\begin{document}

\section{Introduction and summary \label{intro}}

Defining the thermodynamic entropy of a black hole in a quantum theory of gravity -- even in principle -- is an interesting open problem. Any candidate formula should take into account the quantum
fluctuations of matter and gravity and reduce to the Bekenstein-Hawking formula in the
classical, large horizon area limit. Ultraviolet fluctuations can be incorporated
by applying Wald's generalization of the Bekenstein-Hawking formula 
using the Wilsonian quantum effective action, but infrared fluctuations
need an extra prescription.  Such a proposal, called the {\it quantum entropy function},
has been put forward recently \cite{Sen:2008vm} for the case of charged extremal 
black holes in any number of dimensions.

This proposal relies on the near-horizon geometry of an extremal black hole 
being $AdS_2$ (or rather, a patch of global $AdS_2$ known as the $AdS_2$ 
black hole \cite{Maldacena:1998bw,Spradlin:1999bn}) times a compact manifold $M$. 
The  quantum entropy function
$d(q_i)$ is a Euclidean path integral over asymptotically $AdS_{2}$ field configurations with fixed 
electric charge $q_i$, fixed value of the scalar fields
at infinity, and a Wilson line insertion. The functional integral runs over 
all fields in the dimensionally reduced  two-dimensional field theory. 
The proposal comes with a specific prescription for 
dealing with the infrared divergence due to the infinite volume of $AdS$ space. The ultraviolet divergences 
of Einstein gravity  in $d \ge 3$ are assumed to be resolved by some 
ultraviolet completion such as string theory; they translate into the existence of an infinite number of massive fields in
two dimensions, irrelevant at low energies. 

The Euclidean 
path integral is dominated by the field configuration corresponding to pure $AdS_2$, 
but there are in general other saddle points  approaching  $AdS_2$ asymptotically, and leading 
to exponentially suppressed contributions. These saddle points
do not necessarily correspond to smooth geometries, 
but may include (e.g. orbifold) singularities  allowed by the UV completion. A proposal for 
including such orbifolds has been made in  \cite{Banerjee:2008ky,Sen:2009vz}. 

In this work, we explain and refine this construction in the 
case where the $AdS_2$ black hole is the 
``very near horizon'' limit \cite{Maldacena:1998bw,Spradlin:1999bn} of a  
BTZ black hole in $AdS_3$, 
which could itself be embedded in a larger  asymptotically flat space.
We  further test this understanding in the case of $\cN=4$ dyons in 4 dimensions,
where the exact degeneracies are captured by a certain Siegel modular form.
We compute the contributions to the dyon degeneracies from arbitrary 
poles in the Siegel upper half plane, and find agreement  with the classical action of the 
$SL(2,\IZ)$ family of $AdS_3$ black holes. Finally, we relate the sum over poles 
to the Farey tail expansion of a certain Jacobi form, which should arise as
the modified elliptic genus of the superconformal field theory 
dual to $\cN=4$ dyons. 

The existence of an intermediate $AdS_3$ region requires that  
the black hole arises as a black string winding around an extra 
circle. This includes most of the examples in string theory where the
microscopic origin of the black hole entropy has been understood \cite{Strominger:1996sh,
Maldacena:1997de}. In this case
the geometry which dominates the Euclidean path integral 
asymptotes to $AdS_2\times S_1 \times \tilde M$ with $\tilde M$ a compact manifold. 
The circle $S_1$ is non-trivially fibered over  $AdS_2$ so
as to produce a constant field strength after  Kaluza-Klein reduction.
An infinite family of saddle points, 
labelled by two relatively prime integers 
$(c,d)$ with $1 \le d < c $ can then be constructed as follows: 
consider an $\IZ/c\IZ$ orbifold of the 
dominant saddle point, where the cyclic generator acts as a $2\pi/c$
rotation in Euclidean $AdS_{2}$, accompanied by 
a translation of angle $2\pi d/c$ along the circle $S^{1}$. 
When $c>1$ and $ 1 \le c < d$, the resulting geometry is smooth, and gives
a subleading contribution  of order
\be
\label{expsub}
\exp\left(\frac{\cS_{0}}{c} + 2 \pi \I \, q \,\frac{d}{c} \right)
\ee
to the quantum entropy function, 
where $\cS_{0}$ is the contribution of the dominant  configuration with $(c,d)=(1,0)$,
equal to the Bekenstein-Hawking-Wald macroscopic entropy, and $q$ is  
the momentum on the circle $S^{1}$.

In the language of the parent  $AdS_{3}$, these geometries are the extremal limit\footnote{One has to be careful in taking this limit. The boundary torus of Euclidean $AdS_{3}$ has a complex structure modulus $\tau$.  Upon zooming into the very-near-horizon region of the Euclideanized extremal BTZ black hole, one gets another torus with complex structure $(\tau \to 0,\bar \tau \to \infty)$, which does not
admit any real sections. We discuss this in section \S{\ref{ads2farey}}.} of 
the $\Gamma_\infty \backslash SL(2,\IZ)/ \Gamma_\infty$ family of $AdS_3$ black holes 
discussed in \cite{Maldacena:1998bw,Kraus:2006nb,Manschot:2007ha}. The geometry of thermal $AdS_{3}$  is a solid torus, and the various solutions correspond to all possible ways of  filling in the boundary torus with a three-dimensional smooth manifold of constant negative curvature. The two integers $(c,d)$ label the unique cycle of the boundary torus
which becomes contractible in the bulk. 

In the present case, the mass (or angular momentum) of the BTZ black hole is fixed. 
The family of  $AdS_{3}$  solutions that contribute to the entropy function therefore have asymptotic complex structure varying as a function of $(c,d)$. 
From the point of view of the microscopic theory, 
the $AdS_{3}$ path integral corresponds to the canonical partition function keeping  the electric potential fixed. The $AdS_{2}$ path integral  corresponds to the microcanonical ensemble with fixed electric charge. Going from one to the other involves  summing over states with different charges, or summing over different boundary conditions; the former is the original $AdS_{3}$ Farey tail, and the latter is what we discuss in this paper.

Subleading corrections of order \eqref{expsub} have been encountered in a recent analysis
of the exact microscopic degeneracies of $\cN=4$ dyons \cite{Banerjee:2008ky}.
There is now overwhelming evidence that the latter are  
encoded as Fourier coefficients of certain Siegel modular 
forms \cite{Dijkgraaf:1996it, LopesCardoso:2004xf,Gaiotto:2005hc, David:2006yn, 
Banerjee:2008pu, Dabholkar:2008zy}. The saddle points in the
semi-classical expansion of the Fourier coefficients at large charges
are labelled by five integers $(m_1,m_2,j,n_1,n_2)$, transforming linearly as 
the 5-dimensional representation of $Sp(2,\IZ)$. For $n_2=1$, the 
saddle point contribution $\exp(\cS_0)$ reproduces the Bekenstein-Hawking-Wald entropy
$\cS_0$, including $R^2$-type 
quantum corrections to the four-dimensional low energy effective action.
For $n_2 > 1$, the saddle point contributes a subleading
correction of order \cite{Banerjee:2008ky}
\be
\exp\left[\frac{\cS_{0}}{n_{2}} 
+ \frac{2\pi\I }{n_2}
\left( n_1 \frac{Q^2}{2} - \frac{j}{2} (P\cdot Q) - m^1 \frac{P^2}{2} \right) \right] \ ,
\ee
precisely\footnote{In the $D1-D5-P-KKM$ 
duality frame, the quadratic combinations become $Q^{2}/2 = q, 
P\cdot Q=l, P^2/2= Q_1 Q_5$, 
where $q$ is the momentum along the circle $S^{1}$ discussed above and $l$ is the 
momentum around a different circle inside $\tilde M$.} 
of the form \eqref{expsub}. Moreover, the partition function $\CZ_m(\rho,v)$ at fixed 
magnetic charge $P^2/2$ and fixed potentials $(\rho,v)$ conjugate to
$(Q^2/2,P\cdot Q)$ can be obtained by summing over all poles with $0\leq m^1<n_2$. 
This provides a Poincar\'e series representation of the Jacobi
form $\CZ_m(\rho,v)$ which is very similar to the Farey tail expansion, and hints at 
some intriguing relation between the Fourier coefficients of the elliptic genus of K3 
and those of the Dedekind function. At any rate, this Farey tail-type expansion supports the
existence of an effective
black string description for any charges, and therefore the existence of 
an intermediate $AdS_3$ region.

The rest of this note is organized as follows. In section \ref{qefsec}, 
we review the quantum entropy formalism of \cite{Sen:2008vm}. In section \ref{ads2farey},
focusing on extremal black holes with an intermediate $AdS_3$ region we construct an 
infinite family of solutions which are asymptotic to $AdS_2$, and lift 
to the extremal limit of the $\Gamma_\infty \backslash \Gamma/ \Gamma_\infty$ 
family of $AdS_3$ black holes familiar from the black hole ``Farey tail".
In Section 4, we proceed to analyze the exponentially suppressed corrections
to  degeneracies of $\cN=4$ dyons, and derive a ``Farey tale" representation of 
the black hole partition function as a sum over poles in the Siegel upper half plane.
The mathematically oriented reader may skip directly to Section \ref{dyon}.

\section{Review of the quantum entropy function formalism\label{qefsec}}

The quantum entropy function  \cite{Sen:2008vm}  generalizes the Wald entropy formula 
to include non-local quantum corrections in a consistent quantum theory of gravity such as string theory. 
It is formulated for extremal black holes, whose near horizon geometry is $AdS_{2} \times M$ 
where $M$ is a compact manifold. The higher dimensional theory is written as a
two-dimensional  theory with a generally infinite set of fields, including the 2D metric, gauge fields 
$A^{i}$ with field strengths $F^{i}$ and matter fields $\phi_{a}$ governing the shape of $M$. 
The magnetic 
charges in higher dimension appear as fluxes on $M$, and generate a potential for the scalars
$\phi_a$ as well as theta-angle couplings for the  field strengths $F^{i}$. There are also higher-derivative contributions to the two-dimensional Lagrangian, induced by ultraviolet 
fluctuations in higher dimensions above the Wilsonian cut-off.

The most general near horizon field configuration consistent with the $SL(2,\IR)$ symmetry of $AdS_{2}$ is:
\be\label{nearhor}
\de s^{2} = v \left[ -(r^{2}-1) \de u^{2} + \frac{\de r^{2}}{r^{2} -1}   \right] \ ,\qquad 
F^{i} = e^{i} \de r \wedge \de u\ , \qquad \phi^{a} (u,r)= \phi^{a}_0\ .
\ee
where $v, e^{i}$ and $\phi^{a}_0$ are constants. 
This is the metric of an $AdS_{2}$ black hole  \cite{Maldacena:1998bw,Spradlin:1999bn} 
with horizon at $r = 1$. It is locally isometric to $AdS_2$ and the region $r>1$
covers a triangular wedge extending halfway from the boundary 
into global $AdS_2$ \cite{Spradlin:1999bn}. 

An analytic continuation $u\to- \I u_E$ leads to the Euclidean metric
\be\label{nearhoreucl}
\de s^{2} = v \left[ (r^{2}-1) \de u_E^{2} + \frac{\de r^{2}}{r^{2} -1}  \right]\ , \qquad 
F^i= -\I\, e^{i} \de r \wedge \de u_E, \qquad \phi^{a} (u_E,r)= \phi^{a}_0\ .
\ee
This metric is non-singular at the erstwhile horizon $r^{2}=1$ 
provided  the Euclidean time
coordinate $u_E$ is periodic modulo $2 \pi$. In the gauge $A^i_{r} =0$, the gauge 
fields  are given by  
\be\label{ggefld} 
A^{i} = -\I\, e^{i} (r -1) \de u_E \, ,
\ee
where the constant term ensures that the Wilson line $\oint_{S^1} A^i$ around the 
thermal circle vanishes at the horizon $r=1$. This is needed for regularity since
the thermal circle contracts to zero size. 

The quantum entropy function is defined as  a functional integral over all field
configurations which asymptote to the
$AdS_{2}$ Euclidean black hole \eqref{nearhoreucl}. Specifically, one requires
the fall-off conditions \cite{Castro:2008ms}
\be
\label{asympcond}
\begin{split}
\de s^2_{0} &= v \left[ 
\left(r^2+\cO(1)\right) \de u_E^2+ \frac{\de r^2}{r^2+\cO(1)}  \right]\  ,\\
&\phi^a =\phi^a_0 + \cO(1/r)\ ,\qquad
A^i = -\I\, e^{i} (r -\cO(1) ) \de u_E\ ,
\end{split}
\ee
which are invariant under an action of the Virasoro algebra.
In particular, in contrast to higher dimensional instances of the AdS/CFT correspondence,
the mode of the gauge field corresponding 
to the electric charge  grows linearly (non-normalizable) and must be kept fixed,
while the mode corresponding to the electric potential  is  constant (normalizable), and  
allowed to fluctuate.

The quantum entropy function, a function of the electric charges $q_i$ and
moduli $\phi^a_0$, is then defined as the functional integral
\be\label{qef}
\Omega (q_i,\phi^a_0) = \left\langle \exp[-\I \, q_i \oint_{u_E}  A^i]  \right\rangle_{\rm{AdS}_2}^{finite}\ .
\ee
The superscript refers to the following prescription for regulating the divergence due to  the infinite volume of the $AdS_{2}$: First, one enforces a cutoff at a large $r = r_{0}$ (more general
cut-offs have been recently discussed in \cite{Sen:2009vz}). As 
$r_{0} \to \infty$,  the proper length $L \sim 2 \pi \sqrt{v} r_{0}$ of the boundary 
goes to infinity, and the integrand always scales as $e^{C_1 r_0 + C_0 + \cO(r_0^{-1})}$.
The finite part is defined as $e^{C_0}$. 

In the classical limit, the functional integral \eqref{qef} is dominated by the saddle point 
where all fields take their classical values \eqref{nearhoreucl}. In this case, the path integral reduces to 
\be\label{classlim1}
\left\langle \exp[-\I \, q_i \oint_{u_E}  A^i]  \right\rangle \sim \exp{ \left( - S_{bulk} - S_{bdry} - \I q_{i}  \oint_{u_E}  A^i  \right)}  \ ,
\ee  
where 
\be\label{Avalue}
S_{bulk} = \oint_{u_E} (r_{0} - 1) \, v \,  \CL \, \de u_E
\ee
is the regulated two-dimensional action. 
Since the integrand is independent of $u_{E}$, the integral simply produces a factor of $2 \pi$.  $S_{bdry}$ has a divergent part proportional to $r_{0}$ and no constant part. 
The divergent part can be removed by adding an appropriate counterterm in the boundary action,
leaving
\be\label{classlim0}
\Omega(q_i,\phi^a_0) \sim e^{2 \pi ( q_{i} e^{i} - v \CL)}  \equiv e^{\cS_0}\ ,
\ee  
where the electric field $e^i$ is related to the  charge $q_{i}$ via 
\be
\label{qerel}
q_{i} = \frac{\p (v \CL)}{\p e^{i}} \ .
\ee
As shown in \cite{Sen:2008yk}, 
the classical action $\cS_0$ reproduces the Bekenstein-Hawking-Wald entropy
of the extremal black hole.

Quantum corrections to the classical answer \eqref{classlim0} are of two types: (i) fluctuations around the 
classical field configuration \eqref{nearhoreucl}, which produce power law corrections, and (ii) non-perturbative effects come from different classical solutions with the same asymptotics as  \eqref{nearhoreucl} (and fluctuations about those configurations), which are exponentially
suppressed with respect to \eqref{classlim0}.

In \cite{Banerjee:2008ky,Sen:2009vz}, it was proposed that there is a universal series of non-perturbative corrections to the degeneracy of the form $e^{\cS_{0}/c}$ for $c$ integer coming from orbifolds of $AdS_{2}\times M$.  We shall see in the next section that this is indeed borne out for 
BTZ black holes.

\section{Subleading saddle points for extremal BTZ black holes\label{ads2farey}}

In this section, we construct an infinite family of solutions asymptotic to 
extremal BTZ black holes, and find that they lead to exponentially suppressed
contributions of order \eqref{expsub} to the quantum entropy function.

\subsection{The  BTZ black hole}

The general solution of three-dimensional gravity with scalar curvature $-6/\ell^2$ 
asymptotic to $AdS_3$
is given by the two-parameter family of BTZ black holes  \cite{Banados:1992wn}, 
\be\label{BTZ}
\de s_3^2 = - \frac{(\rho^2 - \rho_+^2)(\rho^2 - \rho_-^2)}{\rho^2} \de t^2 
+ \frac{\ell^2 \rho^2} {(\rho^2 - \rho_+^2)(\rho^2 - \rho_-^2)} \de\rho^2 
+ \rho^2 \left(\de y - \frac{\rho_+ \rho_-}{\rho^2} \de t \right)^2 \ ,
\ee
where  the azimuthal coordinate of $AdS_3$ at infinity $y$
has periodicity $2 \pi$. The parameters $\rho_+>\rho_->0$ denote the location
of the outer and inner horizon, and depend on the mass $M$
and angular momentum $J$  via $M=(\rho_+^2+\rho_-^2)/(8G \ell^2), 
J=\rho_+ \rho_-/(4 G \ell)$; henceforth we set $8G=1$.  


The solution \eqref{BTZ} is well-known to be an orbifold of $AdS_3$ \cite{Banados:1992gq,
Carlip:1995qv}: to see this, define for the exterior region $\rho>\rho_+$, 
\be
z \equiv \sqrt{\frac{\rho_+^2-\rho_-^2}{\rho^2-\rho_-^2}}\, e^{(\rho_+ y - \rho_- t)/\ell}\ ,\quad
w_\pm \equiv  \sqrt{\frac{\rho^2-\rho_+^2}{\rho^2-\rho_-^2}}\, e^{(\rho_+\mp\rho_-)( y \pm t)/\ell}
\ee
The coordinates $(w_+,w_-,z)$ parametrize an element $g$ of $G=SL(2,\IR)$,
 \be
 \label{sl2g}
 g = \begin{pmatrix} 1 & w_+ \\ 0 & 1 \end{pmatrix}\cdot 
 \begin{pmatrix} z & 0 \\ 0 & 1/z \end{pmatrix}\cdot
 \begin{pmatrix} 1 & 0 \\ w_- & 1 \end{pmatrix}
 \ee
 and the metric \eqref{BTZ} is locally the bi-invariant metric on the group manifold $G$,
 $\de s_3^2= \ell^2(\de z^2 + \de w_+ \de w_-)/z^2$. Globally, 
 the periodicity $y\sim y+2\pi$ implies the identification 
 $g \sim g_L \cdot g \cdot g_R$ where
 \be
 g_L =  \begin{pmatrix} e^{\pi(\rho_+-\rho_-)/l} & 0 \\ 0 & e^{-\pi(\rho_+-\rho_-)/\ell} \end{pmatrix}\ ,
 \qquad
 g_R =  \begin{pmatrix} e^{\pi(\rho_+ + \rho_-)/l} & 0 \\ 0 & e^{-\pi(\rho_+ +\rho_-)/\ell} \end{pmatrix}
 \ee
are two hyperbolic elements in $G$.

The Euclidean section is obtained by analytically continuing both the time coordinate
$t \to -\I t_E$ and the parameter $\rho_- \to \I r_-$, and letting $t_E, r_-$ be real.
Regularity of the Euclidean section at $\rho_+$ requires identifying
\be
\label{perT}
(t,y) \sim \left( t+ \frac{\I}{T}, y+\frac{\I \Omega}{T}\right)  \ ,\qquad
T=\frac{\rho_+^2-\rho_-^2}{2\pi \ell \rho_+}\ ,\quad \Omega = \frac{\rho_-}{\rho_+}\ ,
\ee
which amounts to the trivial identification
\be
\label{wth}
(w_+, w_-,z) \sim \left(e^{2\pi\I} w_+, e^{-2\pi\I} w_-, z \right)
\ee
on the group manifold $G$.
The Euclidean section is a  two-dimensional solid torus filled with an hyperbolic metric. 
The A-cycle $(t(s), y(s))=(t_0 + \I s /T,y_0+\I \Omega s/T)$ with $0\leq s<1$ is contractible in the
full geometry, hence identified as the thermal circle, 
while the B-cycle $(t(s), y(s))=(t_0,y_0+ 2\pi s)$ is non-contractible.
The complex structure of the torus $T^2$ generated by $\pa_{t_E}, \pa_y$
at fixed radial distance $\rho$ is parametrized by the modulus 
\be\label{taurhop}
\tau_{+} = \frac{\I}{\ell}\left( \rho_{-} + \rho_+ \sqrt{\frac{\rho^2-\rho_-^2}{\rho^2-\rho_+^2}}\right)\ ,
\ee
We define 
\be\label{taurhom}
\tau_{-} = \frac{\I}{\ell}\left( \rho_{-} - \rho_+ \sqrt{\frac{\rho^2-\rho_-^2}{\rho^2-\rho_+^2}}\right)\ .
\ee
such that $\tau_{+}$ and $\tau_{-}$ are complex conjugate to each other when $\rho_-$ 
(hence the angular momentum) is imaginary.
At large radius, the complex structure of the induced metric on the torus goes to a constant, 
\be\label{tauinf}
\tau^{\infty}_{\pm} =\frac{\I}{\ell}\ ( \rho_{-} \pm \rho_+)\ .
\ee

\subsection{The extremal limit \label{extlim}}
The extremal limit corresponds to $\rho_+\to \rho_-$ or $\ell M \to J$, such that 
the temperature $T$ goes to zero. Before taking the limit, it is convenient to change
coordinates to 
\be
\label{defruphi}
r\equiv \frac{2\rho^2- \rho_+^2-\rho_-^2}{\rho_+^2-\rho_-^2}\ ,\qquad
u\equiv \frac{1}{\ell}(\rho_+-\rho_-)(t+y)\ ,\qquad 
\phi\equiv y - \frac{\rho_-}{\rho_+} t\ ,
\ee
such that the group element \eqref{sl2g} is now parametrized by
\be
z=\sqrt{\frac{2}{r+1}}\, e^{R\phi/2}\ ,\qquad
w_+ = \sqrt{\frac{r-1}{r+1}}\, e^{u}\ ,\qquad
w_- = \sqrt{\frac{r-1}{r+1}}\, e^{R\phi-u}\ ,
\ee
where $R\equiv 2 \rho_+/\ell$. In these coordinates, the metric \eqref{BTZ}
takes the form
\be\label{extrBTZ}
\de s^2_3 = \frac{\ell^2}{4} \left[ -(r^2-1)\de u^2 + \frac{\de r^2}{r^2-1} + 
R^2 \left(\de \phi + \frac{1}{R} (r-1) \de u  \right)^2  \right] \ ,
\ee
while the thermal and angular identifications are, respectively,
\be\label{iden}
(u,\phi)\sim (u+2\pi\I, \phi) \sim \left(u+\frac{2\pi}{\ell}(\rho_+-\rho_-) ,\phi+2\pi \right)\ .
\ee

We now take the extremal limit $\rho_+\to\rho_-$, keeping the coordinates
$(r,u,\phi)$ and parameter $R$ fixed. The metric stays as in
\eqref{extrBTZ}, but the thermal and angular identifications simplify to 
\be\label{extrid}
(u,\phi)\sim (u+2\pi\I, \phi) \sim(u,\phi+2\pi)\ .
\ee
To leading order in $\lambda\equiv (\rho_+ - \rho_-) /2\to 0$, the change of variable
\eqref{defruphi} coincides with the one considered in 
 \cite{Sen:2008yk,Gupta:2008ki}
\be\label{extrlim}
\rho=\rho_+ + \lambda(r-1)\ ,\quad 
t=\frac{\ell}{4\lambda}u\ ,\quad
y=\phi+\frac{\ell}{4\lambda} \left(1-\frac{2\lambda}{\rho_+}\right) u\ .
\ee
In the extremal limit, the complex structure \eqref{taurhop},\eqref{taurhom} of 
the $(u,\phi)$ torus reduces to 
\be\label{taur}
\tau_{\pm} = \frac{\I R}{2}\left( 1\pm \sqrt{\frac{r+1}{r-1}}\right) \, .
\ee
It is a characteristic feature of the near-horizon
geometry that  $\tau_{+}\sim \I R$ goes to a finite value
while $-1/\tau_{-}\sim -2\I r/R$ diverges linearly as $r\to\infty$.  
More invariantly, the left-moving complex structure $\tau_{+}$
is regular at $r=\infty$ while the right-moving one $\tau_{-}$ reaches a cusp of the 
moduli space $\mathcal{H}/SL(2,\IZ)$. This is possible because
$\tau_+$ and $\tau_-$ are not complex conjugate to each other in the Lorentzian
geometry, and $\rho_-$ needs to stay real for the  extremal limit to exist. 

\subsection{A family of extremal solutions}

Given a complex structure on $T^2$ labelled by $\tau^{\infty}$ or one of its images 
$(a\tau^{\infty}+b)/(c\tau^{\infty}+d)$ with $ {\scriptsize  \left(\begin{array}{cc} a & b \\
c & d \\ \end{array} \right)}  \in SL(2,\IZ)$, there exists a unique 
hyperbolic metric with $T^{2}$ as boundary. There is however, an infinite family of physically distinct  smooth solutions differing by the homology class of the contractible cycle in the full
geometry \cite{Maldacena:1998bw,Dijkgraaf:2000fq}.

It can be parameterized by two coprime integers $(c,d)$  as follows: 
given the parameters $\rho_+,\rho_-$ of the original metric \eqref{BTZ},
and the corresponding asymptotic complex structure moduli \eqref{tauinf},
we define transformed parameters $\rho'_+,\rho'_-$ via
\be\label{newpars}
\frac{\I}{\ellp} ( \rho'_{-} \pm \rho'_+) \equiv \frac{1}{d+c/\tau^{\infty}_{\pm}}\ ,
\ee
and  coordinates $\rho',y',t'$ via
\be\label{newvars}
y' \pm t'   \equiv  - \left(b+ a/\tau^{\infty}_{\mp} \right) (y \pm t) \, , \quad \quad  
\frac{{\rho'_+}^2-{\rho'_-}^2}{{\rho'}^2-{\rho'_-}^2}  \equiv  \frac{\rho_+^2-\rho_-^2}{\rho^2-\rho_-^2}. 
\ee
The solution labelled by $(c,d)$ is obtained by replacing all quantities in the BTZ solution \eqref{BTZ} by corresponding primed quantities. Thus, the metric ${\de s'}_3^2= {\ellp}^2
(\de {z'}^2 + \de w'_+ \de w'_-)/{z'}^2$ is still locally isometric to $AdS_3$,  with
\be
\label{transw}
w'_\pm \equiv  \sqrt{\frac{{\rho'}^2-{\rho'}_+^2}{{\rho'}^2-{\rho'}_-^2}}\, e^{(\rho'_+\mp\rho'_-)( y' \pm t')/\ell}
= \sqrt{\frac{{\rho}^2-{\rho}_+^2}{{\rho}^2-{\rho}_-^2}}\, \exp\left(
\pm\I (y\pm t)  \frac{b+a/\tau^{\infty}_{\mp}}{d+c/\tau^{\infty}_{\mp}} \right)\ 
\ee
and similarly for $z'$. 
The original solution \eqref{BTZ} 
is recovered for $(c,d)=(1,0)$, while $(c,d)=(0,1)$ reproduces global $AdS_3$ with $\rho'_+=0$.
All these solutions have the same asymptotics 
\be\label{asymmet}
\de {s'}^{2}_3 = {\rho'}^{2} (- \de {t'}^{2} + \de {y'}^{2}) + {\ellp}^{2}  \frac{\de {\rho'}^{2}}{{\rho'}^{2}} \quad {\rm as} \  \rho' \to \infty \, ,
\ee
and coordinate periodicities  
\be\label{period}
(t',y') \sim \left( t'+ \frac{\I}{T}, y'+\frac{\I \Omega}{T}\right) \sim \left(t',y'+ 2\pi \right) \ ,
\ee
where $T,\Omega$ are the temperature and angular velocity
of the original solution \eqref{perT}.
However they differ in the homology 
of the thermal circle (i.e. the one which is contractible in the bulk).
The latter is obtained by demanding that the argument of $w'_\pm$ in \eqref{transw}
varies by $\pm 2\pi\I$ as in \eqref{wth}, i.e.
\be\label{therm}
(t'(s), y'(s))=\left(t'_0 +  \frac{\I c\,s}{T} ; y'_0+  \frac{\I c\,s\, \Omega}{T}  + 2\pi d \,s\right) 
\ee
with $0\leq s\leq 1$. 
Changing $d\to d+c$ does not affect the contractible cycle, so inequivalent solutions are
labelled by double cosets $\Gamma_\infty \backslash SL(2,\IZ) / \Gamma_\infty$
where $\Gamma_\infty = {\scriptsize \begin{pmatrix} 1 & * \\ 0 & 1 \end{pmatrix}}$.

We now take the extremal limit of these solutions by taking $\rho'_{+} \to \rho'_{-}$ and zooming in 
the region $\rho' \sim \rho'_{+}$. We do this as before by changing coordinates
\be
\label{defruphipr}
r' \equiv \frac{2\rho'^2- {\rho'_+}^2-{\rho'_-}^2}{{\rho'_+}^2-{\rho'_-}^2}\ ,\qquad
u' \equiv \frac{1}{\ellp}(\rho'_+-\rho'_-)(t'+y')\ ,\qquad 
\phi' \equiv y' - \frac{\rho'_-}{\rho'_+} t'\ ,
\ee
and taking the above limits  keeping $(r',u',\phi')$ fixed. 
In these coordinates, the metric becomes 
\be
\label{extrBTZ2}
\de {s'}^2_3 = \frac{{\ellp}^2}{4} \left[ -({r'}^2-1)\de {u'}^2 + \frac{\de {r'}^2}{{r'}^2-1} + 
{R'}^2 \left(\de \phi' + \frac{1}{R'} (r'-1) \de u'  \right)^2  \right] \ ,
\ee
with $R' = 2 \rho'_{+}/\ellp$, the identifications \eqref{period} translate to
\be\label{orbifold}
(u',\phi') \sim (u'+2\pi \I / c, \phi'-2\pi  d/c)  \sim (u', \phi' + 2 \pi) \ ,
\ee
while the thermal circle is independent of $(c,d)$, 
\be\label{thermagain}
(u'(s),\phi'(s)) \sim (u'_0+2\pi \I s, \phi'_0) \ .
\ee
Comparing \eqref{orbifold} with \eqref{extrid},  it is apparent that 
the extremal limit of the solution labelled by $(c,d)$  is a $\IZ/c\IZ$  orbifold of the 
solution \eqref{extrBTZ}, with $R$ replaced by
$R'$, by a translation  
\be\label{orbifoldg}
\gamma_c :\quad (u,\phi) \mapsto (u+2\pi \I / c, \phi-2\pi  d/c)\ .
\ee
Near $r=1$, the Euclidean geometry looks like  $(\IR^2 \times S_1)/\IZ_c$, where 
$\gamma_c$ acts by a $2\pi/c$ rotation around the origin
of the plane times a translation of angle $2\pi d/c$ along the circle $S_1$
parametrized by $\phi$.  Since $(c,d)$ are relatively prime, this action has no fixed point and the quotient is smooth, as must be the case as the original family of solutions was smooth.

We now observe that the family of distinct extremal solutions
given by \eqref{extrBTZ2} with periodicities \eqref{orbifold},
all have the same asymptotics, 
namely $AdS_{2} \times S^{1}$, provided $1/R'=c/R+\I d $ is kept fixed
while varying $(R,c,d)$.
Indeed, in coordinates
$r'=c r, u'=u/c, \phi'=\phi+ \I (d/c) u$,  the metric takes the form
\be\label{BTZnew}
\de {s'}^2_{3} = \frac{{\ellp}^2}{4} \left[ 
-\left({r}^2-\frac{1}{c^2}\right ) \de {u}^2+ \frac{\de {r}^2}{{r}^2-\frac{1}{c^2}} + 
{R'}^2 \left(\de \phi +  \left(r-\frac{1}{c}\right) \frac{\de u}{R'}  - \I \frac{d}{c} \de u \right)^2 \right]\ ,
\ee
with the same coordinate periodicities as in \eqref{extrid}.
It is easy to check that the fall-off conditions \eqref{asympcond} 
with $v={\ellp}^2/4$ are indeed satisfied for any $(c,d)$ coprime, $c\geq 1$. 
As mentioned in the introduction, in contrast to the $AdS_{3}$ Farey tail, 
 the mass (or angular momentum) of the BTZ black hole is fixed, while the 
 complex structure varies as a function of $(c,d)$. This is consistent with the microcanonical ensemble required for the quantum entropy function.

In the presence of fermions and if the original theory was supersymmetric, the extremal 
BTZ solutions admit Killing spinors \cite{Coussaert:1993jp}. These supercurrents depend only on 
the $r$ coordinate in the solution \eqref{extrBTZ2dnew}, and therefore are not affected by the orbifold. 
Thus, the family of extremal solutions that we constructed preserve the same amount of supersymmetry.

\subsection{The quantum entropy function for BTZ black holes}

We can now apply the quantum entropy function formalism of \S{\ref{qefsec}} to the 
extremal BTZ black hole. Since $\phi$ is a compact direction, the three-dimensional 
Einstein action may be reduced to two dimensions using the Kaluza-Klein ansatz
\be\label{KK}
\de s_3^2 = \de s_2^2 + \ell^2\, e^{-2\psi} (\de  \phi + \cA )^2\ ,
\ee
leading to the two-dimensional action \cite{Strominger:1998yg}
\be
\label{S2}
S= \int d^2 x \sqrt{-g} \left[ e^{-\psi} \left( R + \frac{2}{\ell^2} \right) - \frac{\ell^2}{4} e^{-3\psi} F^2 \right]
+ \dots \ ,
\ee
where the ellipses denote contribution from the extra fields in three dimensions, and from Kaluza-Klein
modes. 

In particular, the BTZ metric \eqref{extrBTZ} provides a classical solution
to \eqref{S2} with
\be
\label{extrBTZ2d}
\de s_2^2   =  \frac{\ell^2}{4}  \left[ -(r^2-1)\de u^2 + \frac{\de r^2}{r^2-1} \right] , \qquad 
e^{-2\psi} = \frac{R^2}{4}  \ ,\qquad
\cA =  \frac{1}{R} (r-1) \de u \ .
\ee
The two-dimensional metric $\de s^2_2$ is just the
two-dimensional $AdS_2$ black hole \eqref{nearhor}, 
with  constant electric flux $e =  1/R$.  Similarly,
the family of solutions \eqref{BTZnew} reduces to solutions to
\eqref{S2} with the same asymptotics and charge as (\ref{extrBTZ2d}). This family
of solutions will therefore also contribute to the 
quantum entropy function. 

It is easiest to compute this contribution by regarding  the $(c,d)$ solution as a freely acting orbifold of the solution \eqref{extrBTZ2d}: 
\be\label{extrBTZ2dnew}
\de s_2^2   =  \frac{\ell^2}{4}  \left[ -(r^2-1)\de u^2 + \frac{\de r^2}{r^2-1} \right] , \qquad 
e^{-2\psi} = \frac{R^2}{4}  \ ,\qquad
\cA =  \left( \frac{1}{R} (r-1) +  \I \, d \right)  \de u \ , 
\ee
with $u \sim u + 2 \pi \I /c$. Since the solutions are locally isometric to \eqref{extrBTZ2d}, 
the Lagrangian density in coordinates $(u,r)$ is constant and independent of $(c,d)$. 
In the classical limit \eqref{classlim1}, the $(c,d)$ dependence appears in the periodicity of the $u$ variable and the discrete Wilson line. The contribution of the 
bulk action \eqref{Avalue}  is 
\be\label{newact}
A^{(c,d)}_{bulk}  = \frac{2 \pi}{c} (r_{0} - 1) \, v \,  \CL 
\ee
where $v=\ell^2/4$, and the value of the Wilson line is
\be\label{newWilson}
\I q \oint_{u_E}  \CA = 2 \pi \I \,  \,q \, d/c \ .
\ee
By putting a cut-off in the radial direction at $r=r_{0}$ in \eqref{extrBTZ2dnew}
(equivalently $r=r_0/c$ in \eqref{BTZnew}) and discarding 
the linearly divergent part, one finds that the solution labelled by
 $(c,d)$ contributes in the classical limit to the quantum entropy function \eqref{qef} as
\be\label{newsolact}
\Omega^{(c,d)}(q) = \exp\left( \frac{\cS_{0}}{c} + 2 \pi \I\, q \,\frac{d}{c} \right) \ ,
\ee
where  $\cS_{0}$ is the contribution for $(c,d)=(1,0)$, i.e. the Wald entropy. Thus,
the family of solutions with $c>1$ leads to  a series of exponentially suppressed corrections
of the form \eqref{newsolact}. 

Our conclusion agrees in spirit with the proposal put forward in  
\cite{Banerjee:2008ky, Sen:2009vz}, but yields a more precise identification of the 
orbifold action in the case of BTZ black holes. In particular, the geometry associated 
to the subleading saddle points appears to be smooth, and the orbifold acts trivially
on the compact manifold $\tilde M$. In the next section, we discuss the case of 
dyons in $\cN=4$ string backgrounds in more detail, where the compact manifold
$M=S_1 \times S_1 \times S_2 \times K3$ allows for more general choices of the
orbifold action.

\section{The dyon partition function in $\CN=4$ theories \label{dyon}}

In this section, we study the exponentially subleading contributions to the degeneracies
of dyons in $\CN=4$ string vacua, and find agreement with the general structure 
found in Section \ref{ads2farey}. We also develop a general ``Farey tale'' expansion 
for the partition function of $\cN=4$ dyons at fixed value of the magnetic charge $P^2/2$,
and contrast it with the usual ``Farey tail'' series governing the $AdS_{3}$ partition function.

\subsection{The degeneracy formula}

We first summarize some well-known facts about dyon degeneracies in $\cN=4$
string backgrounds, referring e.g. to \cite{Sen:2007qy} for more details. While our 
construction can be 
easily extended to other $\cN=4$ backgrounds, we focus for simplicity on 
the heterotic string compactified on a six-dimensional torus $T^6$, or equivalently
type II string on $K3 \times T^2$. The resulting four-dimensional theory is invariant under 
S and T-duality,
\begin{equation}
\label{dualitygroup}
G(\mathbb{Z}) \equiv SL(2, \mathbb{Z})  \times O(22, 6; \mathbb{Z}) \ .
\end{equation}
Dyons are labelled by their electric and magnetic charges $(Q^i,P^i), i = 1, \dots, 28$,
transforming linearly as a $({\bf 2}, {\bf 28})$ representation of $G(\IZ)$. Both $Q^i$
and $P^i$ take values in an even self-dual lattice $\Lambda$ of signature (22,6),
the Narain lattice of the heterotic torus. The automorphism group
of $\Lambda$ defines
the discrete subgroup $O(22, 6; \mathbb{Z})\subset O(22, 6,\IR)$.
The orbits of $(Q^i,P^i)$ under $O(22,6; \mathbb{Z})$ are labelled by 
the quadratic combinations $Q^2/2$, $P^2/2$, and $P \cdot Q$,  invariant
under the continuous T-duality, and the discrete invariant $I = \gcd(Q^i P^j - Q^j P^i) \in \IZ^{+}$,
which is also invariant under S-duality. 
All dyons in the same orbit  carry the same  indexed degeneracy 
$\Omega(Q^2/2, P\cdot Q, P^2/2, I)$\footnote{For brevity we omit the dependence
of $\Omega$ on the values of the moduli at spatial infinity, and correspondingly
the ambiguity in the choice of integration contour in \eqref{inverse3}. The resulting
ambiguities in $\Omega$ scale like $\exp(Q)$ and are still much smaller than the 
exponentially suppressed corrections of interest for this paper.}.
Here we restrict to the simplest case $I=1$, referring to \cite{Banerjee:2008pu, 
Dabholkar:2008zy,Sen:2009vz} for generalizations.

The  indexed degeneracies $\Omega(Q^2/2, P\cdot Q, P^2/2,  I=1)$
can be packaged into a partition function 
$\CZ(\rho, v,\sigma)$, a function of  three complex variables $(\rho, v, \sigma)$
acting as chemical potentials for the  T-duality invariants 
$(Q^2/2, P\cdot Q, P^2/2)$, respectively. As first conjectured in \cite{Dijkgraaf:1996it}, 
$\CZ$ is a Siegel modular form of weight $k=-10$, i.e.  it
satisfies 
\begin{equation}\label{phi}
   \CZ [(A \tau + B )(C\tau + D ) ^{-1}] =  \left[ \det{(C\tau + D )}\right]^k\, 
   \CZ (\tau)
\end{equation}
for $k=10$, where $\tau = {\scriptsize  \left(\begin{array}{cc} \rho & v \\
v & \sigma \\ \end{array} \right)}$ parametrizes Siegel's upper half-plane 
\begin{equation}
\label{cond1}
   \Im \rho > 0, \qquad \Im \sigma > 0, \qquad
   (\Im \rho)(\Im\sigma) > (\Im v)^2\ ,
\end{equation}
and  $g= {\scriptsize  \left(\begin{array}{cc} A & B \\
C & D \\ \end{array} \right)}$ is any element of  $Sp(2, \mathbb{Z})$, i.e.
any integer valued matrix such that $g J g^t =J$ where 
$J= {\scriptsize  \left(\begin{array}{cc} 0 & -{\bf 1} \\
{\bf 1} & 0 \\ \end{array} \right)}$:
\begin{equation}
\label{cond}
   AB^T=BA^T, \qquad  CD^T=DC^T, \qquad AD^T-BC^T= \mathbf{1}\, .
\end{equation}
More specifically, for the heterotic string compactified on $T^6$, $\CZ$ is the inverse of 
Igusa's cusp form $\Phi_{10}$,  which is the unique cusp form of weight $-k=10$ 
under  $Sp(2, \mathbb{Z})$:
\begin{equation}\label{igusa}
    \CZ= \frac{1}{\Phi_{10}} \ .
\end{equation}
Alternatively, $\Phi_{10}$ can be obtained as the square of the product of all even genus 2 theta functions, or as the additive lift of the index 1, weight 10 Jacobi form $\eta^{18}(\rho)\, \theta_1^2(\rho,v)$, or as the multiplicative lift of the elliptic genus of K3. The latter characterization means
that \cite{Gritsenko:1996-3}
\be
\CZ = \frac{\exp\left( \sum_{m=1}^{\infty} e^{2\pi\I m \sigma}\, V_m \cdot \chi_{K3} \right)}{\eta^{18}(\rho)\, \theta_1^2(\rho,v)}
\ee
where
\be
\label{ellK3}
 \chi_{K3}(\rho,v) = 24 \left( \frac{\theta_3(\rho,v)}{\theta_3(\rho,0)}
                           \right)^2
     -2 \frac{ \left[ \theta_4^4(\rho,0)-\theta_2^4(\rho,0) \right]
          \theta_1^2(\rho,v)}{\eta^6(\rho) }
\ee
and $V_m$ are Hecke operators, acting on the Fourier coefficients $c(N,l)$
of a Jacobi form $\phi$ of weight $k$ via \cite{Eichler:1985ja} 
\be
\label{hecke}
V_m \cdot \phi (\rho,v) = \sum_{N,l} \left( \sum_{d|(N,l,m)} d^{k-1}\, 
c( N l/d^2, l/d) \right) \, e^{2\pi\I(N \rho+l v)} \ .
\ee

Given the dyon partition function $\CZ$, the indexed degeneracies can be found from  
\begin{equation}\label{inverse3}
   \Omega(P,Q) = (-1)^{P\cdot Q +1}\int_{\CC}
   \de\rho\, \de v\,  \de\sigma  \, e^{-i\pi ( Q^2 \rho +2 P\cdot Q v + P^2 \sigma)}\,
   {\CZ(\rho,v,\sigma)}
\end{equation}
where (for an appropriate choice of moduli at spatial infinity) the contour $\CC$
is given by 
\be
0 < {\Re(\rho)} \leq 1, \quad 0 < \Re(v ) \leq 1\ ,\quad 0 < \Re(\sigma) \leq 1\ ,  
\ee
while $\Im(\rho),   \Im(v ), \Im(\sigma)$ are fixed to some large positive value.
In this framework, S-duality is realized as an $SL(2,\IZ)$ subgroup of $Sp(2,\IZ)$\ ,
under which $(\rho, v, \sigma)$ transforms as a three-vector dual to 
$(P^2/2,P\cdot Q,Q^2/2)$:
 \be
 \label{Sdual}
g_S = \left(
\begin{array}{llll}
 a & -b & 0 & 0 \\
 -c & d & 0 & 0 \\
 0 & 0 & d & c \\
 0 & 0 & b & a
\end{array}
\right) \ ,\qquad
\begin{pmatrix} \rho' \\ v' \\ \sigma' \end{pmatrix} =
\begin{pmatrix} a^2 & -2 ab & b^2 \\ 
-ac & ad+bc & -bd  \\ c^2 & -2cd & d^2 \end{pmatrix}
\begin{pmatrix} \rho \\ v \\ \sigma \end{pmatrix}\ .
\ee
Invariance of $\CZ$ under $SL(2,\IZ)\subset Sp(2,\IZ)$ ensures that the right-hand side of \eqref{inverse3}
is invariant under S-duality. The reason for covariance 
under the full Siegel modular group is less clear, except in the context of string network
constructions \cite{Gaiotto:2005hc,Dabholkar:2006bj}.

Rather than extracting the Fourier coefficients of $\CZ(\rho,v,\sigma)$ with respect to its three
arguments, it is useful to consider the partition function $\CZ_m(\rho,v)$ for black holes
at fixed values of $m=P^2/2$, but arbitrary values of $Q^2/2$ and $P\cdot Q$: 
\be
\label{Zmdef}
\CZ_m(\rho,v) =  
\int_{0+\I \Im(\sigma)}^{1+\I\Im(\sigma)} d\sigma \, \CZ(\rho,v,\sigma) \, e^{-2\pi\I m\sigma}\ ,
\ee
where again $\Im(\sigma)$ is kept fixed and large. The modular property 
\eqref{phi} for the subgroup $SL(2,\IR)_\rho \ltimes H_3$ of $Sp(2,\IZ)$
of matrices of the form 
\be
\label{sl2rhoSp}
{\scriptsize \begin{pmatrix} a & b\\c &d\end{pmatrix}}_\rho
=\left(
\begin{array}{llll}
 a & 0 & b & 0 \\
 0 & 1 & 0 & 0 \\
 c & 0 & d & 0 \\
 0 & 0 & 0 & 1
\end{array}
\right)\ :\quad
(\rho', v', \sigma') =
\left( \frac{a \rho+b }{c \rho+d },\frac{v}{c \rho+d},\sigma -\frac{c v^2}{c \rho+d }\right)\
\ee
and
\be
\label{HeisSp}
\tilde T_{\lambda,\mu,\kappa}=    \left(
\begin{array}{llll}
 1 & 0 & 0 & \mu  \\
 \lambda  & 1 & \mu  & \kappa \\
 0 & 0 & 1 & -\lambda  \\
 0 & 0 & 0 & 1
\end{array}
\right)\ :\quad\\
\begin{pmatrix}
\rho' \\ v' \\ \sigma'
\end{pmatrix}
 = \begin{pmatrix}\rho \\
 v+ \mu +\lambda  \rho\\ 
 \sigma +\kappa +2\lambda  v+\lambda  \mu  +\lambda^2  \rho
   \end{pmatrix}\ ,
\ee
implies that $\CZ_m(\rho,v)$ is a Jacobi form 
of weight $k=-10$ and index $m$, i.e. it satisfies\footnote{These relations may fail if one crosses
poles in the $\sigma$ plane when deforming the contour back to its original location. This does not
happen provided the imaginary part $\Im \sigma_*$ for all poles is bounded from above, and 
the contour in \eqref{Zmdef} is chosen at a sufficiently large value 
of $\Im\sigma$. \label{foobou}} 
\bea
\CZ_m\left( 
\frac{a \rho+b }{c \rho+d },\frac{v}{c \rho+d}\right)
&=& (c \rho+d)^{k} \, e^{2\pi \I\frac{m c v^2}{c \rho+d }}\, \CZ_m(\rho,v) \ ,\\
\CZ_m\left(\rho, v+\lambda\rho+ \mu\right)&=&
 e^{-2\pi\I m [\lambda( \mu +2 v) + \lambda^2 \rho]}\, \CZ_m(\rho,v) \ .
\eea
$\CZ_m(\rho,v)$ is however not a holomorphic function of $v$, since it has a second order pole at 
the theta divisor $v\in \IZ + \rho \IZ$.
This pole cancels in the product 
\be
\label{eta18Z}
\CZ_m^{5D}(\rho,v) =  \eta^{18}(\rho)\, \theta_1^2(\rho,v) \, \CZ_m(\rho,v)\ ,
\ee
which is a holomorphic Jacobi form of weight $0$ and index $m+1$.
Physically, $\CZ_m^{5D}(\rho,v)$ is the elliptic genus of the D1-D5 superconformal
field theory, counting 5D black hole microstates with $Q_1 Q_5=m$, momentum
$n$ and angular momentum $l$,  
\be
\CZ_m^{5D}(\rho,v) = 
\sum_{n,l} \Omega^{5D}(Q_1 Q_5, n, l)\, e^{2\pi\I( n \rho + l v)} \ .
\ee
The equation \eqref{eta18Z} can be used to systematically evaluate the asymptotic expansion of the 5D black hole degeneracy \cite{Castro:2008ys, Banerjee:2008ag}. Our interest will be on the
meromorphic partition function $\CZ_m(\rho,v)$, which should correspond to the elliptic genus
of the SCFT dual to $\cN=4$ dyons.

\subsection{Mapping the poles}

Our aim will be to evaluate the contour integrals \eqref{inverse3} and \eqref{Zmdef} by 
use of Cauchy's residue formula. In this subsection, we describe the pole structure 
of the partition function $\CZ$, and find an explicit $Sp(2,\IZ)$ transformation
which maps any of them to the standard diagonal divisor $v=0$.

The partition function $\CZ$ is well known to have a second order 
pole\footnote{Our construction straightforwardly generalizes to Siegel modular forms 
of arbitrary weight $k$, with a pole
of arbitrary order at $v=0$, or to modular forms invariant under a finite index subgroup
of the Siegel modular group.}  at the diagonal divisor $v=0$, where it behaves as 
\be
\label{resv0}
\CZ(\rho,v,\sigma) = \frac{1}{v^2\, g(\rho,\sigma)} + \cO(v^0)\ ,
\ee
where 
\be
g(\rho,\sigma)=\eta^{24}(\rho) \, \eta^{24}(\sigma)
\ee
is a Hilbert modular form\footnote{For our purposes, a Hilbert modular form 
of weight $w$ is a function of $\rho,\sigma$ which is a modular form of weight $w$ 
in each argument, and invariant under the exchange $\rho\leftrightarrow\sigma$,
see e.g. \cite{MR2409678}.} of weight $2-k$.
By $Sp(2,\IZ)$ invariance, $\CZ$ must have a second order pole at all 
images of the diagonal divisor, i.e. at the quadratic divisors
\be
\label{divdef}
D(m^i,j,n_i; \Omega) \equiv m^2 - m^1 \rho + n_1 \sigma + n_2 ( \rho\sigma- v^2) + j v = 0\ ,
\ee
where $j$ is any odd integer and  the 5 integers $M=(m^1,m^2,j,n_1, n_2)$ 
are constrained to satisfy
\be
\label{deldef}
\Delta(M) \equiv  j^2 + 4 (m^1 n_1 + m^2 n_2)=1\ .
\ee
The diagonal divisor $v=0$ corresponds to $M=(0,0,1,0,0)$, with $\Delta(M)=1$.
The union of all quadratic divisors \eqref{divdef} with $\Delta(M)=1$ defines the
first Humbert surface \cite{MR930101}. The invariance of the  constraint \eqref{deldef} can be made
manifest by fitting  $M$ into an anti-symmetric anti-traceless bilinear form in $\IC^4$,
\be
M = \begin{pmatrix}
0 & -m^2 & \frac{j}{2} & n_1 \\
 m^2 & 0 & m^1 & -\frac{j}{2} \\
 -\frac{j}{2} & -m^1 & 0 & -n_2 \\
 -n_1 & \frac{j}{2} & n_2 & 0 \end{pmatrix}\ ,
\ee
such that $M$ transforms as $M' = \Omega M \Omega^t$ and 
$\Delta(M)=4\Pf(M)$ is manifestly invariant. This realizes the local isomorphism
$Sp(2)=SO(2,3)$. Moreover, one may check that under a simultaneous $Sp(2,\IZ)$
action on $M$ and $\Omega$, \eqref{divdef}  transforms with weight $-1$,
\be
\label{transD}
D(M'; \Omega') =  [\det(C\Omega+D)]^{-1}\, D(M,\Omega)\ .
\ee

It will be important to determine the residue of $\CZ$ on the general quadratic divisor \eqref{divdef}.
For this purpose, it suffices to find a $Sp(2,\IZ)$ transformation which maps $M_1=(0,0,1,0,0)$
to an arbitrary $M=(m^1,m^2,j,n_1, n_2)$ satisfying 
\be
\label{cons1}
m^1 n_1 + m^2 n_2 = \frac{1-j^2}{4}\ .
\ee
Moreover, we shall insist
that the choice of this transformation is covariant with respect to $SL(2,\IZ)_\rho$.
We shall restrict our attention to $(n_1,n_2)\neq (0,0)$.
It is then useful to choose coprime integers $(k_1,k_2)$ such that
\be
k_2 n_1-k_1 n_2=r \ ,
\ee 
where $r$ is the greatest common divisor of $(n_1,n_2)=r(n_1',n_2')$.
When  \eqref{cons1} is obeyed, $r$ must  divide $(1-j^2)/4$. The solutions
to \eqref{cons1} can then be  parametrized as 
\be
\label{solm}
m^1 = - \frac{j^2-1}{4r} k_2 + \alpha n_2'\ ,\qquad m^2 = \frac{j^2-1}{4r} k_1 - \alpha n_1'\ ,\qquad
\ee
where both $\alpha n_1'$ and $\alpha n_2'$ must be integer. Since $(n_1',n_2')$
are coprime, this amounts to requiring that $\alpha$ is integer. 
Note that $(k_1,k_2)$ are defined up to the addition of an integer multiple of $(n_1',n_2')$:
this can be reabsorbed into a shift of $\alpha$ by an integer multiple of 
$(j^2-1)/4r$, which is integer. We further define 
\be
\delta\equiv \alpha \mod r\ .
\ee
 Since $r|(j^2-1)/4$, we may further decompose  $r=r_1 r_2$ into a product 
of relatively prime factors, where $r_1$ divides $(j+1)/2$ and $r_2$ divides $(j-1)/2$:
\be
\label{j12}
j+1 = 2 r_1 j_2\ ,\qquad j-1 = 2 r_2 j_1\ ,\qquad r_1 j_2 - r_2 j_1 = 1\ .
\ee
The most general solution is given by
\be
\label{js12}
j_1 = s_1 + r_1 L\ ,\qquad j_2= s_2 + r_2 L\ ,\qquad
j = 2 r L + j_0
\ee
where $s_1,s_2$ are fixed integers with $r_1 s_2 - r_2 s_1=1$, $L$ is an arbitrary integer,
and $j_0\equiv r_1 s_2 + r_2 s_1$.

Having defined these number theoretic quantities, it is now easy to check that 
\be
\label{hgen}
h = {\scriptsize \begin{pmatrix} 1 & (\alpha-\delta)/r \\ 0 &1 \end{pmatrix}}_\sigma \cdot  
{\scriptsize \begin{pmatrix} -k_1 & -n_1/r\\ k_2 &n_2/r \end{pmatrix}}_\rho \cdot  
{\scriptsize \begin{pmatrix} 
j_2 & 0 & 0 & j_1 \\
 \delta j_2  & j_2 & j_1 &  \delta j_1 \\
r_2 m_1 & r_2 & r_1 & -r_1 m_1 \\
r_2 & 0 & 0 & r_1\end{pmatrix}}
\ee
is an element of $Sp(2,\IZ)$ mapping $M_1=(0,0,1,0,0)$ into $M=(m^1,m^2,j,n_1,n_2)$.
Clearly, $h$ is ambiguous modulo right multiplication by an element in the stabilizer of $M_1$,
i.e. in the Hilbert modular group $SL(2,\IZ)_\rho \times SL(2,\IZ)_\sigma 
\ltimes (\rho\leftrightarrow \sigma)$.
As we discuss shortly, the choice \eqref{hgen}
has the advantage of being covariant with respect
to $SL(2,\IZ)_\rho$.

Denoting by $\tau'=(\rho',v',\sigma')$ the image\footnote{For comparison
with other discussions in the literature, our change of variable 
reduces for $M=(0,0,1,0,1)$, $k_1=-1,k_2=\alpha=\delta=0$ to
\be\nn
\rho' =\frac{\rho}{(v-1)^2-\rho\sigma}\ ,\qquad
v' =\frac{v(1-v)+\rho\sigma}{(v-1)^2-\rho\sigma}\ ,\qquad
\sigma' =\frac{\sigma}{(v-1)^2-\rho\sigma}\ .
\ee
The  change of variable for general $M$ is too cumbersome to be displayed, 
but follows immediately  from $\tau'=(A \tau + B )(C\tau + D )^{-1}$ with 
$h^{-1}= {\scriptsize  \left(\begin{array}{cc} A & B \\
C & D \\ \end{array} \right)}$ the inverse of \eqref{hgen}.}
 of $\tau$ under $h^{-1}$, the 
quadratic divisor  \eqref{divdef} is therefore mapped to $D(M_1,\tau')=0$, i.e. $v'=0$.
For later use, we record the Jacobian from $(\rho,v,\sigma)$ to
$(\rho',v',\sigma')$: it is given by  ${\pa \tau}/{\pa \tau'}= [\det(C \tau +D)]^{3}$
where $\det(C \tau +D)$ equals
\be
\frac{1}{r_1^2}
\det\begin{pmatrix}
\frac{j+1}{2} (k_2\rho+ k_1) + \delta (n_2 \rho+n_1) +r v &
\alpha +\frac{j-1}{2}\delta - v \left( \frac{j+1}{2} k_2 + \delta n_2 \right) - r \sigma \\
n_2\rho+n_1 & - n_2 v + \frac{j+1}{2} \end{pmatrix}\ .
\ee

Having found a suitable $Sp(2,\IZ)$ transformation mapping the general divisor
 \eqref{divdef}  back to the diagonal divisor $v=0$,
it is now straightforward to extract the residue of $\CZ$ 
on  \eqref{divdef}  using \eqref{resv0} in the primed coordinates. 
For example, expanding around 
$\sigma=\sigma_*$ where $\sigma_*$ is the location of the pole in the $\sigma$ plane,
\be
\label{sigmast}
\sigma_* = \frac{m^1\rho-m^2+ n_2 v^2 -j v}{n_2\rho+n_1}\ , 
\ee
we have 
\be
\begin{split}
\rho' &= -\frac{1}{\rho_0} \left( 1 - {r_2^2}
(\sigma- \sigma_*)/{\rho_0}  + \dots \right)\ ,\\
v' &= -\frac{r}{\rho_0} (\sigma- \sigma_*)  \left( 1 - {r_2^2}(\sigma- \sigma_*)/{\rho_0} 
+ \dots \right)\ ,\\
\sigma' &= \sigma_0 + r_1^2 (\sigma- \sigma_*) + \dots\ ,
\end{split}
\ee
\be
\det(C \tau +D) =  -  \rho_0\,  (n_2' \rho+n_1') \,   \left( 1 +{r_2^2}
(\sigma- \sigma_*)/{\rho_0}  + \dots \right)\ ,
\ee
where $(\rho_0,\sigma_0)$ are the values of $(-1/\rho',\sigma')$ 
at $\sigma=\sigma_*$, namely
\be
\begin{split}
\label{defrho0}
\rho_0 &= 
\frac{j+1}{2r_1^2} \frac{k_2\rho+k_1}{n_2'\rho+n_1'} +
 \frac{r_2}{r_1} \frac{v}{n_2'\rho+n_1'}+ \frac{r_2}{r_1}\, \delta\  ,\quad \\
\sigma_0 &=  
-\frac{j-1}{2r_2^2} \frac{k_2\rho+k_1}{n_2'\rho+n_1'}
 -\frac{r_1}{r_2} \frac{v}{n_2'\rho+n_1'} +\frac{r_1}{r_2}\, \delta  \ .
 \end{split}
\ee
Using the modular properties of $\CZ(\rho,v,\sigma)$ 
and $g(\rho,\sigma)$, we conclude that on the quadratic divisor \eqref{divdef}, 
\be
\label{Zas}
\begin{split}
\CZ (\rho,v,\sigma) &\sim \frac{ (\rho')^{2-k} [\det(C \tau +D)]^{-k}}
{(v')^2 \, g(-1/\rho',\sigma')} \\
&= \frac{(n_2' \rho+n_1')^{-k}}{r^2(\sigma- \sigma_*)^2 \, g(\rho_0,\sigma_0)} 
\left[ 1 - \frac{ (r_2^2  \pa_{\rho_0} + r_1^2 \pa_{\sigma_0}) 
g(\rho_0,\sigma_0)}{g(\rho_0,\sigma_0)}(\sigma- \sigma_*) 
 + \dots \right] \ .
\end{split}
\ee
It is important to note that the poles \eqref{sigmast} are bounded from above in the $\sigma$
plane: indeed, expressing $m^1$ in terms of $m^2,j,n_1,n_2$ using \eqref{deldef} and
extremizing with respect to $j$, one obtains
\be
\Im \sigma_* \leq \frac{\Im \rho}{4 | n_2 \rho + n_1|^2} + \frac{(\Im v)^2}{\Im\rho}\ ,
\ee
where the upper bound would be reached at $j/2 = n_2 \Re(v) - (n_1 + n_2 \Re\rho) \Im v/\Im \rho$.
A further extremization with respect to $n_1,n_2$ leads to
\be
\label{boundsig}
\Im \sigma_* \leq \frac14 \max(\Im \rho, 1/ \Im \rho) + \frac{(\Im v)^2}{\Im\rho}\ ,
\ee
which ensures that the Fourier coefficient \eqref{Zmdef} is indeed a Jacobi form,
as discussed in footnote \ref{foobou}.

Alternatively, one may expand in the $v$ plane around either of the two roots 
of \eqref{divdef}
\be
\label{defvpm}
v_\pm = \frac{1}{2n_2} \left( j \pm \sqrt{
(n_1 \sigma - m^1 \rho + m^2)^2 - 4 j n_2} \right)\ .
\ee
The asymptotic expansion is obtained by replacing 
\be
\sigma-\sigma_* =\frac{n_2}{n_2\rho + n_1} \left(  \mp (v_+-v_-) (v-v_\pm) + (v-v_\pm)^2 +\dots \right)
\ee
and $v=v_\pm+ (v-v_\pm)$ in the  expansions above.

It is important that our choice of $Sp(2,\IZ)$ transformation is covariant under $SL(2,\IZ)_\rho$: 
if $(\rho,v)$ transform as in \eqref{sl2rhoSp} and if $M$ transforms as 
\be
\begin{pmatrix} n_1 & k_1 & m^2  \\ n_2 & k_2 & -m^1 \end{pmatrix} \mapsto 
\begin{pmatrix} a & -b   \\ -c & d \end{pmatrix}
\begin{pmatrix} n_1 & k_1  \\ n_2 & k_2 \end{pmatrix}\ ,\qquad
(j,\alpha,\delta) \mapsto (j,\alpha,\delta)\ ,
\ee 
then $\rho_0,\sigma_0$ are invariant, while $\sigma_*$ transforms
in the same way as $\sigma$ in \eqref{sl2rhoSp}.
On the other hand, under the spectral flow \eqref{HeisSp}, $n_1,n_2,k_1,k_2$ 
can be taken to be invariant, so that
\be
\begin{split}
&\rho_0 \mapsto \rho_0 -\frac{r_2}{r_1} (k_1 \lambda - k_2 \mu)\ ,\qquad
\sigma_0 \mapsto \sigma_0 + \frac{r_1}{r_2}(k_1 \lambda - k_2 \mu)\ ,\qquad\\
&\qquad \qquad \sigma_*\mapsto \sigma_* + \lambda ( \mu+\lambda \rho+2 v)+ \kappa\ ,
\end{split}
\ee
while $\alpha$ shifts by an integer,
\be
\alpha 
\mapsto \alpha + \frac{(j-\lambda n_1+\mu n_2)[k_2 (\mu n_2-\lambda n_1)+ r \lambda]}
{n_2}+ \kappa r \ .
\ee
In particular,  the spectral flow $\tilde T_{0,0,\kappa}$
takes $\alpha\mapsto\alpha+\kappa r$, and leaves $\delta$ invariant.

\subsection{Subleading contributions to the 4D degeneracies \label{4dsub}}

We now evaluate the integral \eqref{inverse3} by first performing the $\sigma$
integral using Cauchy's residue formula, and evaluating the remaining 
integral over $(\rho,v)$ in the saddle point approximation\footnote{The 
more standard approach where the integral over $v$ is done by the residue theorem
while the ones over $(\rho,\sigma)$ are done in the saddle point approximation
is discussed in Appendix \ref{4dsubf}.}. This approximation becomes exact in the
limit where $Q^2/2, P^2/2, P\cdot Q$ are scaled to infinity keeping their ratio fixed. 
Moreover, we assume that the quartic invariant $P^2 Q^2 - (P\cdot Q)^2$ is positive, 
such that the entropy is dominated by a large dyonic black hole.

The poles that contribute to the $\sigma$ integral
must belong to the strip $0\leq \sigma\leq 1$. Since any pole can be mapped
into this strip by a spectral flow $\tilde T_{0,0,\kappa}$, which maps $(m^1,m^2)\mapsto
(m^1+\kappa n_2,m^2 - \kappa n_1)$, and since the residue is invariant under this action,
we must restrict to poles with $0\leq m^1< n^2$, subject to the quadratic constraint
\eqref{deldef}. Using \eqref{Zas}, we find
\be
\label{Om1}
\begin{split}
  \Omega(P,Q)= & \Omega^{(0)}(P,Q) + (-1)^{P\cdot Q +1}\int_{0}^{1} 
   \de\rho\, \int_{0}^{1}  \de v\, \sum_{(n_1,n_2,j,m^1,m^2), \Delta(M)=1} \\
&    \frac{(n_2' \rho+n_1')^{-k}}{r^2\, g(\rho_0,\sigma_0)} 
\left( \I \pi P^2 + 
 \frac{ (r_2^2  \pa_{\rho_0} + r_1^2 \pa_{\sigma_0}) g(\rho_0,\sigma_0)}{g(\rho_0,\sigma_0)}
\right)   
      \, e^{-\I \pi ( Q^2 \rho +2 P\cdot Q v + P^2 \sigma_*)}\ ,
      \end{split}
\ee
where $\Omega^{(0)}(P,Q)$ includes the contribution with poles with $(n_1,n_2)=(0,0)$,
possibly together with  additional boundary contributions which remain after the 
contour has been deformed across all poles with  $(n_1,n_2)\neq (0,0)$. 
As mentioned below \eqref{Zmdef}, the integrand is a Jacobi form $\CZ_m(\rho,v)$ 
of index $m=P^2/2$ and weight $k=-10$, which we discuss in its own right in the next subsection. 
For now, we proceed with the integral over $\rho$ and $v$.

To leading order in the charges, the integral \eqref{Om1} can be approximated by extremizing
the exponent with respect to $(\rho,v)$, including the fluctuation determinant and evaluating
the prefactor at the saddle point. The saddle point lies at 
\be
\label{sad}
\rho_* = -\frac{n_1}{n_2}+ \I \frac{P^2}{2n_2 \sqrt{P^2 Q^2 - (P\cdot Q)^2}}\ ,\qquad
v_* = \frac{j}{2n_2}+ \I \frac{P\cdot Q}{2n_2 \sqrt{P^2 Q^2 - (P\cdot Q)^2}}\ ,\qquad
\ee
at which point the location of the pole \eqref{sigmast} evaluates to
\be
\sigma_* = \frac{m^1}{n_2}+ \I \frac{Q^2}{2n_2 \sqrt{P^2 Q^2 - (P\cdot Q)^2}}\ .
\ee
Note in particular that $(\rho_*,v_*,\sigma_*)$ transforms as a triplet under S-duality
\eqref{Sdual}. For the saddle point to lie in the integration domain, we require that 
$-n_2<n_1\leq 0$, $1\leq j < 2 n_2$. The value of the exponent at the saddle point 
is given by \cite{Banerjee:2008ky}
\be
\label{Sn4}
\cS_\star = \frac{\pi}{n_2} \sqrt{P^2 Q^2 - (P\cdot Q)^2} + \frac{\I\pi}{n_2}
( n_1 Q^2 - j (P\cdot Q) - m^1 P^2 )
\ee
Thus, as already noted in \cite{Dijkgraaf:1996it}, the poles with $n_2>1$ give
exponentially suppressed contributions with respect to the one with $n_2=1$.
Our interest is in further analyzing the contributions of all the subleading saddle points,
extending the analysis in \cite{Banerjee:2008ky}.

The fluctuation determinant around the saddle point \eqref{sad} is given by
\be
\det\begin{pmatrix} \pa_{\rho^2} \cS & \pa_{\rho v} \cS \\ \pa_{\rho v} \cS &  \pa_{v^2} \cS \end{pmatrix}
= \left( \frac{4\pi n_2 [P^2 Q^2 - (P\cdot Q)^2]}{P^2} \right)^2\ .
\ee
Moreover, the arguments $(\rho_0, \sigma_0)$ of the prefactor in \eqref{Om1}
reduce to
\be
\label{sapt}
\begin{split}
\rho_0^* &= \frac{k_2(j+1)}{2r_1^2 n_2'}
+ \frac{1}{r_1 n_2'\, P^2} \left( 
 - P\cdot Q + \I \sqrt{P^2 Q^2 - (P\cdot Q)^2}\right) \ ,\\
\sigma_0^* &= -\frac{k_2(j-1)}{2r_2^2 n_2'} 
+ \frac{1}{r_2 n_2'\, P^2} \left( 
 P\cdot Q + \I \sqrt{P^2 Q^2 - (P\cdot Q)^2}\right) \ .
\end{split}
\ee
Thus, the saddle point labelled by $(n_1,n_2,j,m^1,m^2)$ with $j$ odd and
\be
\Delta(M)=1\ ,\qquad 0\leq m^1 < n_2\ ,\quad -n_2<n_1\leq 0, \quad 1\leq  j < 2 n_2 \ ,
\ee
contributes to the degeneracies of $\cN=4$ in the semi-classical limit as 
 \be
 \label{subln4}
\frac{(-1)^{P\cdot Q}\, (P^2)^{1-k}}{r^{2-k} n_2 [P^2 Q^2 - (P\cdot Q)^2]^{1-\frac{k}{2}}\, g(\rho_0^*,\sigma_0^*)}
\left( \I \pi P^2  + 
 \frac{ (r_2^2  \pa_{\rho_0^*} + r_1^2 \pa_{\sigma_0^*}) 
g(\rho_0^*,\sigma_0^*)}{g(\rho_0^*,\sigma_0^*)}
\right)
\, e^{\cS_\star}
\ee

When $n_2=|j|=1$ and therefore $n_1=m^1=m^2=k_2=\alpha=\delta=0, r=-k_1=1$, the 
exponent $\cS_*$ in \eqref{Sn4} reproduces the Bekenstein-Hawking entropy of the dyonic black hole.
The prefactor in \eqref{subln4} 
leads to logarithmically and power suppressed corrections  to the entropy
which are consistent with the contributions of the $R^2$-type quantum corrections
to the Wald entropy \cite{LopesCardoso:2004xf}. In fact, the values of 
$(\rho_0, \sigma_0)$ in \eqref{sapt} are equal to the attractor values of the 
axio-dilaton  $(S,-\bar S)$, and the functions $g_1, g_2$ are precisely the ones which 
govern the $R^2$ corrections to the four-dimensional 
effective action \cite{Harvey:1996ir,Gregori:1997hi}. This agreement continues 
to hold off-shell \cite{LopesCardoso:2004xf}, as outlined in Appendix~B.

Instead, contributions with $n_2>1$ are exponentially suppressed compared to those 
with $n_2=1$. They agree qualitatively with the semi-classical contributions of
subleading $AdS_2$ saddle points \eqref{newsolact} computed 
in Section \ref{ads2farey} (strictly speaking for $n_1,n_2$ coprime only).
While \eqref{newsolact} involved only one discrete quantum number $d$ conjugate
to a single charge $q$, \eqref{Sn4} displays the contributions of three discrete quantum
numbers $n_1, -j/2, -m^1$ conjugate to $Q^2/2, P\cdot Q, P^2/2$, respectively.
In the duality frame of the $D1-D5-P-KKM$ system on $S^1\times S^1\times K3$
with one unit of KK monopole charge,
the quadratic invariants  $Q^{2}/2 = q, P\cdot Q=l$ become equal 
to the momenta along the two circles, and so $n_1$ and $j/2$ may be viewed
as the timelike component of the Kaluza-Klein gauge fields $g_{\mu5}$ 
and $g_{\mu6}$ along the two circles. The third combination $m^1 P^2/2= m^1 Q_1 Q_5$
for either $Q_1=1$ or $Q_5=1$ may be interpreted as a discrete Wilson line for 
the Ramond-Ramond one-form or five-form. When $n_1,(j\pm 1)/2, m_1$ and $n_2$ are not
relatively prime, the orbifold described in Section \ref{ads2farey} is no longer freely
acting, but nevertheless the saddle point action still retains the same form \eqref{Sn4}.

In principle, it should also be possible to interpret the prefactor in \eqref{subln4} as the effect
of $R^2$-type corrections around the subleading semi-classical geometry.
For $(n_1,n_2)$ coprime, and therefore $r_1=r_2=1$, the values of  
the ``axio-dilaton" $(\rho_0, \sigma_0)$ appearing as the argument of
the Hilbert modular function $g$ are rescaled by a factor $1/n_2$ 
compared to the values at the leading saddle point. 
If the heterotic coupling is weak in the attractor region, one may 
approximate $\ln g(\rho_0^*,\sigma_0^*)\sim 2\pi\I (\rho_0^*+\sigma_0^*)$,
which is reduced by a factor $1/n_2$ from its value at $(n_1,n_2)=(0,1)$.
This appears to be consistent with our identification of the subleading
saddle point as a $\IZ/n_2 \IZ$ orbifold of the dominant solution. 
More generally, this is consistent with the fact that the effective volume of
 the two-torus $S_1\times S_1$
is effectively reduced by a factor $n_2$ at the ``very near horizon'' $r=1$.
We do not know how to interpret  the additional shift  proportional to $k_2$,
which breaks the reality relation $\rho_0 = - \overline{\sigma_0}$. 
We also explain in Appendix B that the
off-shell agreement with the entropy function which was observed in \cite{LopesCardoso:2004xf}
for the leading saddle point does not seem to extend to subleading saddle points.

\subsection{Poincar\'e series representation\label{poinca}}

We now consider the partition function \eqref{Zmdef} at fixed $m=P^2/2$.
As explained below \eqref{Zmdef}, this is a meromorphic Jacobi form of
weight $k=-10$ and index $m$, which should be identified as the elliptic genus
of the SCFT dual to 4D dyons. In the same way as in Section \ref{4dsub}, 
the integral over $\sigma$ can be evaluated by Cauchy's
residue formula, leading to 
\be
\label{Zmsum}
\begin{split}
\CZ_m(\rho,v)&= \sum
\frac{2\pi\I  (n_2' \rho + n_1')^{-k}}{r^2 g(\rho_0,\sigma_0)} 
\left( 2\pi\I m + 
 \frac{ (r_2^2  \pa_{\rho_0} + r_1^2 \pa_{\sigma_0}) g(\rho_0,\sigma_0)}{g(\rho_0,\sigma_0)}\right) \\
& \exp\left[ -2\pi\I m \left( \frac{m^1\rho-m^2}{n_2\rho+n_1}+ \frac{n_2 v^2 -j v}{n_2\rho+n_1}
\right) \right]
\end{split}
\ee
where the sum runs over all integers 
$j,n_1,n_2,k_1,k_2,\delta,r_1,r_2$ (the condition $0\leq m^1<m^2$ fixes $\alpha$
uniquely, but its precise value is irrelevant). Fourier expanding the prefactor, 
\be
\label{Fourpre}
\frac{2\pi\I m  g(\rho_0,\sigma_0)+   (r_2^2  \pa_{\rho_0} + r_1^2 \pa_{\sigma_0}) g(\rho_0,\sigma_0)}
{r^2\, (g(\rho_0,\sigma_0))^2} = \sum_{N_1, N_2} 
c_{m,r_1,r_2} (N_1,N_2) \, e^{2\pi\I (N_1 \rho_0+N_2 \sigma_0)}
\ee
we obtain
\be
\CZ_m(\rho,v) = 2\pi\I  \, \sum_{j,n_1,n_2,k_1,k_2,r_1,r_2,N_1,N_2,\delta}   (n_2' \rho + n_1')^{-k} 
\, c_{m,r_1,r_2} (N_1,N_2)  \  e^{\cS}
\ee
where the exponent may be written, up to an additive integer, as
\be
\label{Seq}
\frac{\cS}{2\pi\I} = -\frac{m n'_2 v^2}{n'_2\rho+n'_1} + \frac{l v}{n'_2\rho+n'_1}
+ \frac{k_2\rho+k_1}{n'_2\rho+n'_1}
\left(\tilde N+ \frac{l^2}{4m}\right)+ \left( r_2^2 N_1 + r_1^2 N_2- m\right) \frac{\delta}{r} \ ,
\ee
where we have defined
\be
l = \frac{jm}{r} + r \left( \frac{N_1}{r_1^2} - \frac{N_2}{r_2^2} \right)\ ,\quad
\tilde N=\frac1{2}\left(\frac{N_1}{r_1^2} + \frac{N_2}{r_2^2}-\frac{m}{2r^2}\right) - \frac{(N_1 r_2^2-N_2 r_1^2)^2}{4m r^2} 
\ee
The sum over $\delta$ ranging from 0 to $r-1$ vanishes unless $r_2^2 N_1+r_1^2 N_2-m$ is divisible
by $r$, in which case it produces an overall factor of $r$. To solve the congruence, let us choose 
integers $t_1,t_2$ such that 
\be
\label{m12}
m=r_1^2 t_2 + r_2^2 t_1 \ . 
\ee
Since $r_1$ and $r_2$ are coprime, they
must divide $N_1-t_1$ and $N_2-t_2$, respectively: 
\be
N_1 = t_1+ r_1 N_1'\ ,\qquad N_2 = t_2 + r_2 N_2'\  ,
\ee 
where $N_1', N_2'$ are integers. Moreover, using \eqref{j12} and \eqref{m12} , we can write
\be
l=\mu + 2 m L\ ,\qquad \mu\equiv r_2 t_1 \frac{j_0+1}{r_1} - r_1 t_2 \frac{1-j_0}{r_2} 
+ N_1' r_2- N_2' r_1\ ,
\ee
where $\mu$ is manifestly integer. Having defined $\mu$ in this way, 
one may further compute  $N\equiv \tilde N + \frac{\mu^2}{4m}$,
\be
N= \frac{1+j_0}{2r_1} N_1' + \frac{1-j_0}{2r_2} N_2' +
\frac{(1+j_0)^2}{4r_1^2} t_1 + \frac{(1-j_0)^2}{4r_2^2} t_2\ ,
\ee
which is also manifestly integer.

The sum over $L$  produces a unary theta series \eqref{thetadef}
evaluated at $\rho'=(k_2\rho+k_1)/(n'_2\rho+n'_1), v'=v/(n'_2\rho+n'_1)$.
Therefore, identifying
\be
a= k_2,\quad b=k_1\ ,\quad c=n'_2\ ,\quad d=n'_1\ ,
\ee
we recognize the sum over poles \eqref{Zmsum} as a Poincar\'e series,
\be
\label{modfar2}
\phi_m 
=\phi_{m}^{(0)} + \frac12 \sum_{\mu=0}^{2m-1} \, \sum_{\gamma \in \Gamma_\infty\backslash\Gamma} 
(c\rho+d)^{-k}\, e^{-\frac{2\I\pi m c v^2}{c\rho+d}}\, 
h_\mu\left( \frac{a\rho+b}{c\rho+d}\right) \, 
\theta_{m,\mu} \left( \frac{a\rho+b}{c\rho+d}, \frac{v}{c\rho+d}\right)   \ , 
\ee
where $\phi_{m}^{(0)}$ denotes the contributions of the poles with $(n_1,n_2)=0$, and
\be
\label{hmu4}
h_{\mu}(\rho) = 
2\pi\I \, r \, \sum_{r_1,r_2,N_1',N_2'}
c_{m,r_1,r_2}(r_1 N_1'+t_1, r_2 N_2'+t_2) \ 
e^{2\pi\I \left( N -  \frac{\mu^2}{4m} 
\right) \rho}
\ee
where the sum runs over integers with a fixed value of $\mu$ mod $2m$.
Finally, the original Siegel modular form $\CZ$ may be recovered by resumming
the Fourier series,
\be
\label{farsieg}
\CZ = \CZ^{(0)} +  \frac12 \sum_m
\sum_{\mu=0}^{2m-1} \, \sum_{\gamma \in \Gamma_\infty\backslash\Gamma} 
(c\rho+d)^{-k}\, e^{2\pi\I m \left( \sigma-\frac{c v^2}{c\rho+d}\right)}\, 
h_\mu\left( \frac{a\rho+b}{c\rho+d}\right) \, 
\theta_{m,\mu} \left( \frac{a\rho+b}{c\rho+d}, \frac{v}{c\rho+d}\right) 
\ee
To summarize, we have rewritten the sum over the five integers $M=(m^1,m^2,j,n_1,n_2)$ 
modulo the constraint $\Delta(M)=1$ into a sum over cosets 
$\scriptsize \begin{pmatrix} k_2 & k_1 \\ n_2/r & n_1/r \end{pmatrix} \in
\Gamma_\infty\backslash\Gamma$, spectral flow $l=\mu \mod 2m$, and 
quantum numbers $r_1,r_2,N_1',N_2'$. The auxiliary integers $s_1,s_2,t_1,t_2$
are fixed in terms of $m,r_1,r_2$ by $r_1 s_2 - r_2 s_1=1$ and \eqref{m12}.

The Poincar\'e series \eqref{modfar2} is our ``Farey tale" expansion for the $\cN=4$ dyon
partition function. It closely resembles the usual Farey tail expansion \eqref{modfar} 
for Jacobi modular forms. While the sum \eqref{modfar2} is not 
restricted to states satisfying the  ``cosmic censorship" bound $\tilde N<0$, this
restriction may be enforced by hand at no cost, since non-polar
terms (after properly regulating the sum) average to zero \cite{Manschot:2007ha}.

In detail however, the structure of \eqref{modfar2} is considerably more intricate
than \eqref{modfar}. In particular, the Fourier
coefficients of the vector-valued modular form $h_\mu(\rho)$ must take the special
form \eqref{hmu4}, where the coefficients $c_{m,r_1,r_2}(N_1,N_2)$ originate from
expanding \eqref{Fourpre}. This structure ensures that the resulting sum \eqref{farsieg}
is (at least formally) a Siegel modular form of weight $k$, for any choice of a Hilbert modular form
$g(\rho,\sigma)$ of weight $2-k$. In fact, the series \eqref{farsieg} is a Poincar\'e-type series for the Siegel modular group, where the sum runs over 
the coset  $\Gamma_1\backslash Sp(2,\IZ)$, where $\Gamma_1$ is the
stabilizer of $M_1=(0,0,1,0,0)$, i.e. the Hilbert modular group. Thus, it should provide
a lift from Hilbert modular forms to Siegel modular forms with a pole on the diagonal 
divisor $v=0$.

Unfortunately, in contrast to \eqref{modfar}, our expansion \eqref{modfar2} is formal, as we
have not attempted to regulate the sum over poles. In particular, we have little control 
over the ``degenerate contribution'' $\CZ^{(0)}$, which is the part that remains once the contour 
in the $\sigma$ plane has been passed through all the poles with $\Im\sigma_*>0$. This 
can in principle be determined from the ``non-degenerate" contributions $(n_1,n_2)\neq 0$
by requiring that  \eqref{farsieg} is  $Sp(2,\IZ)$ invariant. 
This degenerate contribution should reproduce the expected pole \eqref{resv0} at $v=0$,
together with its images at $v=m^1\rho-m^2$. Thus, it is natural to expect that it is given
by the sum 
\be
\sum_{(m^1,m^2)\in \IZ^2} \frac{1}{(v-m^1\rho+m^2)^2\, g(\rho,\sigma - m^1(m^1\rho-m^2))}\ ,
\ee
while $\phi_{m}^{(0)}$ will be given by the Fourier coefficients of this sum with respect to $\sigma$.
It is an interesting mathematical problem to  turn our ``Farey tale" into a precise mathematical 
statement, and see whether additional constraints must be imposed 
on $g(\rho,\sigma)$ to avoid possible modular anomalies, in the spirit
of \cite{Manschot:2007ha}.

Assuming that the Farey tale expansion can be made rigorous,  it gives an alternative
representation of the Fourier-Jacobi coefficients of $1/\Phi_{10}$ in terms of the 
Fourier coefficients of the Dedekind modular form,
as opposed to the standard Farey tail representation \eqref{modfar} where
$h_\mu$ are directly related to the coefficients of the elliptic genus of $K3$ \eqref{ellK3} 
via the action of the 
Hecke operators \eqref{hecke}. The agreement between the two representations implies
identities between these Fourier coefficients which would be interesting to spell out.

From the physics point of view, it should be possible to give a detailed macroscopic 
interpretation of \eqref{modfar2} as a sum over $AdS_2$ geometries. We have offered
an interpretation of the exponent $\cS_*$ at the saddle point, but clearly more work 
remains to interpret the prefactor especially when $n_1,n_2$ are not relatively prime.
Moreover, it would also be desirable to improve our understanding of the degenerate contributions,
as they play crucial role for consistency with wall-crossing \cite{Sen:2007vb}.
Interestingly, this is tied with the fact that the decomposition \eqref{thetadec} for
meromorphic forms involves vector-valued ``mock"  modular forms rather than usual   
modular forms \cite{Zwegers:2002,Zagier:2007,MockPaper}. 
Hopefully, resolving these issues will shed some light on the microscopic description 
of $\cN=4$ dyons.

\section*{Acknowledgements}

We are grateful to A.~Castro, A.~Dabholkar, S.~Minwalla, E.~Verlinde and A.~Sen  for discussions.
The research of B.P. is supported in part by ANR (CNRS-USAR)
contract no.05-BLAN-0079-01.
The research of S.M. is supported in part by the European Commision Marie Curie Fellowship 
under the contract PIIF-GA-2008-220899.  S.M. would like to thank TIFR, Mumbai for hospitality where part of this work was carried out.

\appendix

\section{Review of the Farey tail expansion}

For the reader's convenience, we briefly summarize the Farey tail expansion of 
a weak holomorphic 
Jacobi form $\phi(\rho,v)$ of weight $k\leq 0$ and index $m$ \cite{Dijkgraaf:2000fq,
Manschot:2007ha}.  Let $c(N,l)$ be the Fourier coefficients of $\phi$,
\be
\label{fourjac}
\phi(\rho,v) = \sum_{N\geq 0, l\in\IZ} \, c(N,l)\, e^{2\pi\I(N \rho+l v)}\ .
\ee
Using spectral flow invariance, $\phi$ can be decomposed as 
\be
\label{thetadec}
\phi(\rho,v) = \sum_{\mu=0}^{2m-1} h_\mu(\rho)\, \theta_{m,\mu} (\rho,v)\ ,
\ee
where $\theta_{m,\mu}$ is an index $m$ unary theta series,
\be
\label{thetadef}
\theta_{m,\mu}(\rho,v) = \sum_{l\in\IZ, \quad l=\mu \mod 2m} 
e^{2\pi\I \left( \frac  {l^2}{4m}\rho+ l v \right)}\ ,
\ee
and $h_\mu(\rho)$ is a vector valued modular form 
\be
\label{hmu}
h_\mu(\rho) = \sum_{N\in \IZ} \, 
c_\mu(4m N-\mu^2) \, e^{2\pi\I \left( N-\frac{\mu^2}{4m}\right) \rho}
\ee
where $c_\mu(4mN-m^2)\equiv  (-1)^{2ml} c(N,l)$ for $l=\mu \mod 2m$.
Note that the holomorphy of $\phi$ with respect to $v$ is essential: 
if $\phi$ has poles in  the $v$-plane, $h_\mu(\rho)$ are only 
``mock" modular forms, and \eqref{thetadec} has to
be supplemented an extra term \cite{Zwegers:2002}.

The Farey tail expansion of  $\phi$ can be obtained by replacing $h_\mu(\rho)$ 
in \eqref{thetadec}  by its Rademacher expansion. In this way one obtains 
 the Poincar\'e series representation
\be
\label{modfar}
\begin{split}
\phi_m &=\frac12 h_\mu\left( \frac{\mu^2}{4m}\right)\, \theta_\mu(\rho,v)
+ \frac12 \sum_{\mu=0}^{2m-1} \, \sum_{\gamma \in \Gamma_\infty\backslash\Gamma} 
(c\rho+d)^{-k}\, e^{-\frac{2\I\pi m c v^2}{c\rho+d}}\, \\
& \qquad \times h_\mu^-\left( \frac{a\rho+b}{c\rho+d}\right) \, 
\theta_{m,\mu} \left( \frac{a\rho+b}{c\rho+d}, \frac{v}{c\rho+d}\right)  
\end{split}
\ee
where  $h^-_\mu$ is the regularized polar part,
\be
h^-_\mu(\rho) = \sum_{N;4mN-\mu^2<0} \, 
c_\mu(4m N-\mu^2) \, e^{2\pi\I \left( N-\frac{\mu^2}{4m}\right) \rho}\,
 R\left( \frac{2\pi\I |N-\frac{\mu^2|}{4m}}{c(c\rho+d)} \right)
\ee
and $R(x)$ is a regularizing factor, such that $R(x)\to 1$ exponentially fast at $x\to\infty$.
The sum runs over $\scriptsize\gamma=\begin{pmatrix} a & b\\c & d\end{pmatrix}$
where $(c,d)=1$ and $(a,b)$ is any one of the solutions of $ad-bc=1$. It may be
regularized by restricting to $|c|\leq K, |d|\leq K$ and letting $K\to\infty$ at the end.
Note that due to the regularization, the Poincar\' e sum for arbitrary choices of the polar 
coefficients $c_\mu$ 
and multiplier system may not be modular invariant. It is possible however to supplement it
with a non-holomorphic term so as to restore modular invariance \cite{Manschot:2007ha}.

\section{Subleading contributions to the entropy function  \label{4dsubf}}

In this appendix, we  evaluate \eqref{inverse3} by a more standard 
procedure \cite{LopesCardoso:2004xf,David:2006yn}, which is to
first perform the integral over $v$ using Cauchy residue's formula, and then treat 
the integral over $\rho,\sigma$ by saddle point
methods. The advantage is that, keeping only the term with  $n_2=1$ in the first step, 
and identifing $(\rho,\sigma)$ with the axio-dilaton 
$S=S_1+\I S_2$ according to
via 
\be
\rho = \frac{\I}{2S_2}\  ,\qquad \sigma = \frac{\I(S_1^2+S_2^2)}{2S_2}\ ,
\ee
the integrand may be recognized as the exponential of the 
macroscopic entropy function, with contributions from $R^2$-type corrections
to the 4D low energy effective action \cite{Sen:2005wa}.

This off-shell agreement between the microscopic partition function and the macroscopic geometry
is quite remarkable. However, it does not seem to extend to exponentially 
suppressed corrections with $n_2>1$. 
Indeed, at the pole $v=v_+$ in \eqref{defvpm},  
the exponent in \eqref{inverse3} becomes
\be
\cS 
= \frac{\pi}{2S_2} | Q- \tau P|^2 - \I\pi  \nu\, P\cdot Q \ ,
\ee
with 
\be
\label{eqnu}
\nu = \frac{j}{n_2} + \frac{\I S_1}{S_2} 
- \frac{1}{n_2 S_2} \sqrt{S_2^2-(n_2-2\I n_1 S_2)(n_2(S_1^2+S_2^2)+2\I m_1 S_2)}\ ,
\ee
while the arguments of the Hilbert modular form $g(\rho_0,\sigma_0)$ 
appearing in \eqref{Zas} are given by
\be
\label{rhopm}
\begin{split}
\rho_0 &= \frac{(1+j)r_2 k_2}{2r_1 n_2}  
+ \frac{\I r_2^2}{n_2(n_2-2\I n_1 S_2)}\left( 
S_2 +\sqrt{S_2^2-(n_2-2\I n_1 S_2)(n_2(S_1^2+S_2^2)+2\I m_1 S_2)} \right)\\
\sigma_0 &= \frac{(1-j)r_1 k_2}{2r_2n_2} 
+ \frac{\I r_1^2}{n_2(n_2-2\I n_1 S_2)}\left( 
S_2 -\sqrt{S_2^2-(n_2-2\I n_1 S_2)(n_2(S_1^2+S_2^2)+2\I m_1 S_2)} \right)
\end{split}
\ee
Thus, the integrand in the $\rho,\sigma$ integral can be written as $e^{-F}$   where
\be
- F =  \frac{\pi}{2S_2} | Q- \tau P|^2  - \I\pi P\cdot Q \, (\nu-1) 
-\log g(\rho_0,\sigma_0) + \dots 
\ee
where the ellipses stand for other contributions to the residue at $v=v_\pm$.
For the leading order contribution  with $(n_1,n_2)=(0,1)$, \eqref{eqnu} and \eqref{rhopm} reduce to
\be
 \nu=1\ ,\qquad \rho_0 = -S_1 + \I S_2\ ,\qquad \sigma_0 = S_1 + \I S_2\ ,\qquad
\ee
and $F$ is recognized
as  entropy function of the four-dimensional black hole, 
including the effect of the $R^2$-type corrections with depence on the 
axio-dilaton $S$ \cite{LopesCardoso:2004xf,David:2006yn}.
For $n_2>1$, this interpretation of the $(\rho,\sigma)$ integral
seems to break down.


\providecommand{\href}[2]{#2}\begingroup\raggedright\endgroup

\end{document}